\documentclass[runningheads]{llncs}
\usepackage[T1]{fontenc}
\usepackage{amsmath}
\usepackage[braket, qm]{qcircuit}
\usepackage[compat=0.6]{yquant}
\usepackage{graphicx}
\usepackage{hyperref}
\usepackage{booktabs}
\usepackage{multirow}
\usepackage{hyphenat}
\begin{document}
\title{Toward Fault-Tolerant Variational Optimization: QAOA under [[4,2,2]] Error Detection}
%
%
\author{Matteo Robert Child\inst{1}\orcidID{0009-0006-6649-5415} \and
Emanuele Dri\inst{2}\orcidID{0000-0002-5144-1514} \and
Giacomo Vitali\inst{2}\orcidID{0000-0002-3056-796X} \and
Chiara Vercellino\inst{2}\orcidID{0000-0002-0562-3157} \and
Alberto Leporati\inst{1}\orcidID{0000-0002-8105-4371}
}
\authorrunning{M.R. Child et al.}
%
\institute{University of Milano-Bicocca, Department of Informatics, Systems and Communication, Italy \and
Fondazione LINKS, 10141 Torino, Italy
}
%
\maketitle              
\begin{abstract}
We present a partially fault-tolerant implementation of QAOA based on the $[[4,2,2]]$ error-detection code, targeting the Max-Cut problem on a square graph. Our main contribution is a novel ancilla-mediated logical $R_{ZZ}$ gate enabling interactions between qubits in different $[[4,2,2]]$ blocks.
We evaluate unencoded and encoded circuits under five noise models, with both all-to-all and grid-routed connectivity, using the Cirq and qsimcirq frameworks with parallel CPU execution. Post-selection on stabilizer measurements consistently improves the probability of sampling optimal bitstrings, with five measurements providing the strongest benefit. These results support error-detection as a practical near-term strategy for improving the quality of variational quantum algorithms.

\end{abstract}
\section{Introduction}
Quantum computing represents a paradigm shift that has the potential to offer novel algorithmic approaches for specific complex problems.
Nevertheless, each of the current physical implementations of the qubit (the fundamental unit of quantum computation) exhibits intrinsic fragility and a pronounced sensitivity to noise. 
This poses a major challenge for the scaling of quantum computers and, consequently, for the algorithms that rely on them.

The scientific community has long identified error-detection and error\hyp{}correction protocols as the solution to this challenge. 
For this reason, the advent of the era of fault-tolerant quantum computation is generating considerable interest, and its robustness to errors will precisely be the result of these protocols.

In addition, it is widely held within the scientific community that the interplay between quantum and high-performance computing is key to unlocking and enabling further avenues in which quantum advantage may be realised.

For these reasons, in this work we propose, build and test a partially fault-tolerant implementation that exploits error-detection protocols for a Max-Cut problem solved using the QAOA algorithm. 
This approach can effectively leverage high-performance computing resources for training, especially when scaling to problem instances of significant size due to parameter concentration properties that allow optimal angles to be pre-computed or optimized in parallel offline.

The remainder of this paper is organised as follows. Section~\ref{sec:background} reviews the two key building blocks of our approach: 
QAOA as a variational algorithm amenable to HPC-assisted training, and the $[[4,2,2]]$ error-detection code. 
Section~\ref{sec:methods} describes the Max-Cut instance, the QAOA circuit construction, and our novel inter-block logical $R_{ZZ}$ gate. 
Section~\ref{sec:results} presents the experimental results across five noise models and two connectivity topologies. 
Section~\ref{sec:conclusions} discusses conclusions and future directions.

\section{Background}
\label{sec:background}
To contextualise the contributions of this work, we review the two key ingredients of our approach: the Quantum Approximate Optimisation Algorithm (QAOA) as a variational method amenable to HPC-assisted training, and the $[[4,2,2]]$ quantum error-detection code that underpins our partially fault-tolerant implementation. 
The Max-Cut problem itself, being well known in the QAOA literature, is introduced directly alongside the circuit construction in Section~\ref{sec:methods}.

Variational Quantum Algorithms (VQAs) have attracted considerable attention in recent years. 
These hybrid approaches rely on parameterised quantum circuits, known as \textit{ansätze}, whose parameters are optimised through a classical outer loop that minimises a cost function evaluated on the quantum device~\cite{cerezo2021variational}. 
A well-known obstruction to the scalability of this paradigm is the barren plateau (BP) phenomenon \cite{mcclean2018barren}: 
as the number of qubits grows, gradients of the cost function concentrate exponentially around zero, so that gradient-based optimisation requires an exponential number of shots. 
A recent line of work has further sharpened the tension between trainability and quantum advantage by showing that, for a broad class of variational models, the very structural features that provably rule out 
barren plateaus also imply that the loss function is classically simulable in polynomial time \cite{cerezo2023simulability}. 

QAOA \cite{Farhi2014QAOA} occupies a comparatively favourable position in this landscape.
Unlike generic VQAs, QAOA targets \emph{bitstring sampling} rather than expectation-value estimation: 
the simulability argument of \cite{cerezo2023simulability} applies to the latter, but does not directly rule out a quantum advantage in the sampling regime \cite{farhiHarrow2016supremacy}.
Moreover, QAOA exhibits \emph{parameter concentration} \cite{Akshay2021,wurtz2023sampling}: optimal angles for a small instance of a structured problem family transfer to much larger instances, drastically reducing the effective training dimensionality and making the optimisation amenable to parallelisation on HPC resources.
Under the large-loop conjecture, the expectation value of the tree subgraph serves as a performance guarantee for any regular graph of the same degree \cite{wurtz2023sampling}, providing analytic grounding that does not rely on gradient-based training.

These features make QAOA particularly well-suited to a quantum--HPC workflow in which classical resources absorb the training cost and the quantum device is used primarily for sampling at fixed parameters. 
Parameter concentration implies that average-case optimal angles can be computed offline on small representative instances and deployed on the quantum device with no, or minimal, retraining; 
this optimisation parallelises trivially across HPC workers and, at low depth, can exploit the lightcone locality of the cost Hamiltonian via tensor-network or subgraph-enumeration methods \cite{wurtz2023sampling}. 
When some on-device refinement is desirable, recent protocols \cite{hao2025endtoend} show that few-shot fine-tuning from a classically-precomputed starting point can be carried out with very limited shot budgets, 
keeping the optimisation burden on the HPC side. Our partially fault-tolerant implementation slots naturally into this workflow: 
the $[[4,2,2]]$ error-detection layer described next is applied on the sampling side, where it most directly improves the quality of the bitstrings returned to the classical optimiser.

The second ingredient of our approach is the $[[4,2,2]]$ quantum error-detection code~\cite{Vaidman1996,Grassl1997}.
Unlike full quantum error correction, which actively recovers from errors, error \emph{detection} relies on post-selection: 
circuit runs in which an error is detected are discarded, yielding a cleaner output distribution at the cost of a reduced sampling rate.
This strategy has recently been demonstrated to improve algorithmic performance for QAOA on hardware, even at moderate system sizes~\cite{hePerformanceQuantumApproximate2025}.
The $[[4,2,2]]$ code encodes two logical qubits into four physical qubits with code distance~2, meaning it can detect (but not correct) any single-qubit error.
Its stabiliser group is generated by $XXXX$ and $ZZZZ$, so a syndrome measurement at the end of a circuit flags any run in which an odd-weight Pauli error has occurred. Flagged runs are then discarded in post-processing---an offline decision made after the fact, in contrast with error \emph{correction}, which applies real-time corrective operations (feedback) to the quantum state before further gates are executed.

The code admits a transversal gate set ($X$, $Z$, $H\otimes H$, and $CNOT$ between different blocks), while non-transversal rotations such as $R_X$ and $R_{ZZ}$ (precisely the gates required by the QAOA ansatz) can be implemented in a \emph{quasi fault-tolerant} manner via an ancilla qubit, as demonstrated in \cite{bedalovFaultTolerantOperationMaterials2024}. These prior works, however, only considered gates acting within a single $[[4,2,2]]$ block.
Our contribution, detailed in Section~\ref{sec:methods}, is an ancilla-mediated implementation of $R_{ZZ}$ gates \emph{between} different blocks, which is the operation required when the two qubits involved in a QAOA cost-layer edge belong to separate logical blocks.

\section{Methods}
\label{sec:methods}

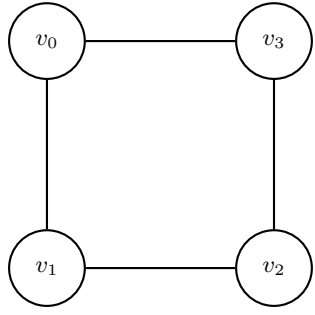
\begin{figure}
\begin{center}
\begin{tikzpicture}[node distance={30mm}, thick, main/.style = {draw, circle, minimum size=1cm}] 
    \node[main] (0) {$v_0$}; 
    \node[main] (1) [right of=0] {$v_3$}; 
    \node[main] (2) [below of=1] {$v_2$}; 
    \node[main] (3) [below of=0] {$v_1$}; 
    
    \draw (0) -- (1);
    \draw (1) -- (2);
    \draw (2) -- (3);
    \draw (3) -- (0);
\end{tikzpicture} 
\end{center}
\caption{Small instance of a \textit{2-regular} square graph with four vertices used as a test case to solve for the Max-Cut problem.}
\label{fig:maxcut}
\end{figure}

The goal of the Max-Cut problem is to partition the vertices of a graph into two sets such that the number of edges between them is maximized. If we map the vertices of a graph into a vector $\mathbf{z} = \{\pm1,\pm1,\pm1, .. \}$ where $0$s and $1$s correspond to the subset to which each vertex belongs, this results in maximizing of the following objective function: $f(\mathbf{z}) = \sum_{(i,j)\in E} \frac{1}{2} (1 - z_i z_j)$. The best known approximation algorithm for the Max-Cut is the Goemans-Williamson algorithm \cite{Goemans1995}, which sets a lower bound to the approximation ratio at $0.878$.
In this work, we focus on the \textit{2-regular} four-vertex cycle graph $C_4$ (Fig.~\ref{fig:maxcut}), whose optimal Max-Cut partitions consist of non-adjacent vertices: the sets $\{v_0, v_2\}$ and $\{v_1, v_3\}$.

Using \textit{little-endian} encoding and the natural vertex-to-qubit assignment $q_i = v_i$, this is represented by the bitstrings 0101 and 1010; note that the encoded implementation in Section~\ref{sec:results} adopts a different qubit assignment, which relabels these solutions as 0110 and 1001.

A promising quantum algorithm for solving NP-hard problems is the QAOA \cite{Farhi2014QAOA}.
QAOA is a discretized adiabatic evolution that aims to evolve the ground state of a known system into the ground state of a target Hamiltonian. This is achieved by an alternating sequence of two operators, the mixer Hamiltonian and the cost Hamiltonian. 

The mixer Hamiltonian is composed of single-qubit rotations:
\begin{equation}
    e^{-i\beta (X_0 + X_1 + X_2 + \dots)} = \prod_{j} e^{-i\beta X_j} = e^{-i\beta X_0} e^{-i\beta X_1} e^{-i\beta X_2} \dots
\end{equation}

while the cost Hamiltonian mirrors the Max-Cut cost function and features a contribution for each edge $(j, k)$ in the graph:
\begin{equation}
    e^{-i\gamma \sum_{\langle j,k \rangle} Z_j Z_k} = \prod_{\langle j,k \rangle} e^{-i\gamma Z_j Z_k} = e^{-i\gamma Z_0 Z_1} e^{-i\gamma Z_1 Z_2} \dots
\end{equation}

In this work, we implemented a logical version of the QAOA using a low logical encoding rate code, since it is less sensitive to the typical noise level of current NISQ devices. Given the previous works mentioned in Section~\ref{sec:background}, we decided to use the $[[4,2,2]]$ code.

The $[[4,2,2]]$ is a hyperbolic self-dual CSS stabilizer code that is the smallest two-logical-qubit stabilizer code to detect a single-qubit error. It encodes two logical qubits into four physical qubits according to the following definitions:

\begin{equation}
\begin{aligned} 
|\overline{00}\rangle &= \frac{1}{\sqrt{2}} \left( \ket{0000} + \ket{1111} \right) \quad |\overline{01}\rangle = \frac{1}{\sqrt{2}} \left( \ket{0011} + \ket{1100} \right) \\ 
|\overline{10}\rangle &= \frac{1}{\sqrt{2}} \left( \ket{0101} + \ket{1010} \right) \quad |\overline{11}\rangle = \frac{1}{\sqrt{2}} \left( \ket{0110} + \ket{1001} \right) 
\end{aligned}
\end{equation}

\begin{figure}
\begin{center}
    \begin{tikzpicture}
        \begin{yquant}
            qubit {$\text{A}_{\idx}$} c0[2];
            qubit {$\text{B}_{\idx}$} c1[2];

            cnot c1[1] | c0[1];
            box {$\mathrm{R_{Z}}(\theta)$} c1[1];
            cnot c1[1] | c0[1];
        \end{yquant}
    \end{tikzpicture}

    \vspace{0.5cm}
    {\huge $=$}
    \vspace{0.5cm}

    \begin{tikzpicture}
        \begin{yquant}
            qubit {$\text{A}_{\idx}$} c1[4];
            qubit {$\text{B}_{\idx}$} c0[4];
            qubit {anc} anc;

            cnot c0[0] | c1[0];
            cnot c0[1] | c1[1];
            cnot c0[2] | c1[2];
            cnot c0[3] | c1[3];
            barrier (c1, c0, anc);

            swap (c0[0], c0[2]);
            cnot anc | c0[1];
            cnot anc | c0[2];
            
            box {$\mathrm{R_Z}(\theta)$} anc;
            cnot anc | c0[0];
            cnot anc | c0[3];
            
            swap (c0[0], c0[2]);
            barrier (c1, c0, anc);
            
            cnot c0[0] | c1[0];
            cnot c0[1] | c1[1];
            cnot c0[2] | c1[2];
            cnot c0[3] | c1[3];
        \end{yquant}
    \end{tikzpicture}
\end{center}
    \caption{Physical implementation of the inter-block $R_{ZZ}$ logical operation using two blocks of $[[4,2,2]]$ code. Logical qubits are identified as $\text{A}_{0,1}$ and $\text{B}_{0,1}$, respectively. Between the two broken barriers is the intra-block procedure from \cite{bedalovFaultTolerantOperationMaterials2024} plus two SWAP gates, acting as logical CNOTs. The SWAP gates can be eliminated by permuting the qubit operands of the CNOTs between them.}
\label{fig:circuito}
\end{figure}
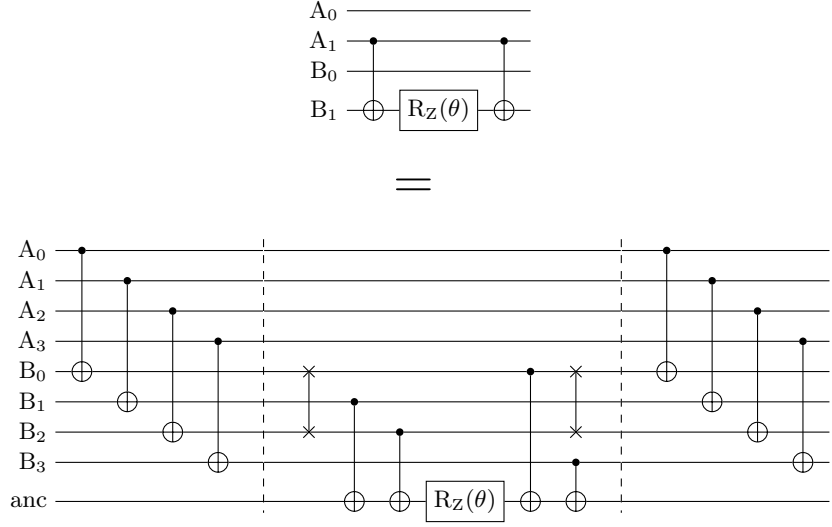

In addition to the known CSS fault-tolerant Clifford gates, in order to implement the logical QAOA we needed to define some non-Clifford operations, in particular $R_{X}$ and $R_{ZZ}$. Since our preferred targets are early fault-tolerant devices, we avoided using magic state distillation techniques, due to their high resource requirement. Instead, we used partially fault-tolerant approaches described in this Section.\\
Logical $R_{X}$ is implemented using the ancilla-mediated procedure of \cite{bedalovFaultTolerantOperationMaterials2024}, with two additional SWAP gates placed according to whether the qubit occupies the upper or lower position in the $[[4,2,2]]$ block. There are three possible $R_{ZZ}$ arrangements. For qubits in the same block, we use the procedure of \cite{bedalovFaultTolerantOperationMaterials2024}, a partially fault-tolerant $R_{ZZ}$ that does not detect Z errors immediately before or after the physical ancilla $R_{Z}$. For qubits in different blocks at the same position (up-up or down-down), we use Fig.~\ref{fig:circuito}; for different positions (up-down or down-up), we supplement it with two logical SWAP gates. This is the main contribution of this work with respect to the algorithm encoding. Because these inter-block logical $R_{ZZ}$ gates combine the intra-block logical $R_{ZZ}$ with transversal fault-tolerant gates, they remain partially fault-tolerant.
Indeed, while more costly than the intra-block logical $R_{ZZ}$, the inter-block operation completes the set of $R_{ZZ}$ primitives: together with the intra-block gate, the same-position and different-position inter-block variants cover every possible arrangement of the two qubits involved in a cost-layer edge. Any Max-Cut instance can therefore be encoded across multiple $[[4,2,2]]$ blocks without defining new physical implementations for the logical $R_{ZZ}$. To detect errors we use the fault-tolerant error detection procedure from \cite{Reichardt2020} to project the state into the code space to detect violations of expected parity without destroying the logical information.

In the next Section we report the results of the experimental campaign performed by emulating the quantum algorithm on classical machines.

\section{Experimental Results}
\label{sec:results}

\subsection{Circuits}

We implemented a two-layer ($p=2$) QAOA circuit and trained its parameters using classical optimization (COBYLA). The optimized parameters were then evaluated on a logical version of the algorithm under various noise models.

Our experiments considered four circuit variants for the square Max-Cut instance. The first is a simple unencoded circuit, used as a reference baseline. The second is a circuit encoded under the $[[4,2,2]]$ code, which exploits reusable ancillas (two ancillas for stabilizer measurements and one ancilla shared between the logical $R_X$ and $R_{ZZ}$ rotations; since these rotations are applied sequentially, the ancilla can be reset and reused between them). The third and fourth variants are, respectively, the unencoded and encoded circuits routed onto Google's Sycamore topology (using cirq's \texttt{RouteCQC} router).
For simplicity, we adopt the assignment in which $v_0$ and $v_1$ are mapped to the upper and lower positions of block~A, while $v_3$ and $v_2$ are mapped to the upper and lower positions of block~B, respectively. In this way, the edges $(v_0, v_1)$ and $(v_2, v_3)$ correspond to intra-block interactions (within blocks A and B, respectively), whereas edges $(v_1, v_2)$ and $(v_0, v_3)$ correspond to direct lower-lower and upper-upper inter-block interactions. Under the little-endian convention $\ket{q_3 q_2 q_1 q_0}$, the two optimal Max-Cut solutions $\{v_0, v_2\}$ and $\{v_1, v_3\}$ are represented by bitstrings $0110$ and $1001$.

\begin{figure}
    \centering
    \includegraphics[width=1.\linewidth]{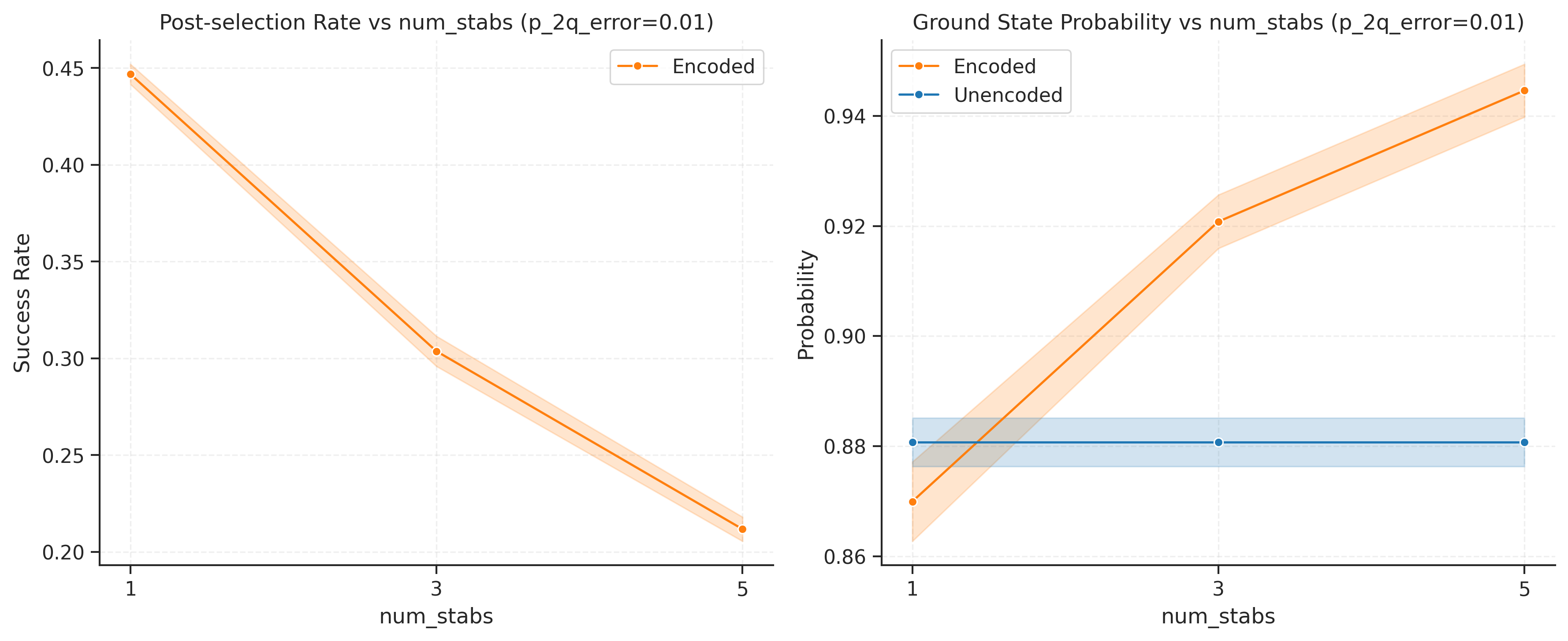}
    \caption{Post-selection rate (left) and ground state probability of post-selected samples (right) for the all-to-all encoded and unencoded circuits under two-qubit depolarizing noise at $p = 0.01$, for three stabilizer measurement counts: only at the end (1); before every $R_{ZZ}$ layer and at the end (3); before every $R_{ZZ}$ layer, every $R_{X}$ layer, and at the end (5). Line colour denotes encoding: orange for the $[[4,2,2]]$-encoded circuit, blue for the unencoded baseline. Shaded bands represent $\pm 1$ standard deviation across 10 experimental runs of 5{,}000 repetitions each. The encoded circuit with 5 stabilizer measurements achieves the greatest improvement in ground state probability.}
    \label{fig:stabilizer-analysis}
\end{figure}

We evaluated the effect of varying the number of stabilizer measurements (1, 3, 5) and found that 5 measurements provided the strongest error protection, as reported in Figure~\ref{fig:stabilizer-analysis}. These consist of one measurement preceding each $R_X$ and $R_{ZZ}$ layer, plus a final measurement that serves as the decoding step and can detect errors whenever the qubits violate the expected parity. Table~\ref{tab:table_depth} summarizes gate counts and the depth of each circuit. The latter is measured as the number of moments.

\begin{table}
    \centering
    \caption{Gate counts and circuit depth for the four circuit variants. The columns report the number of single-qubit gates, two-qubit gates, and depth, where depth is measured as the number of moments (i.e., parallel time steps).}
    \label{tab:table_depth}
    \setlength{\tabcolsep}{8pt}
    \begin{tabular}{@{}llccc@{}}
        \toprule
        \textbf{Encoding} & \textbf{Routing} & \textbf{1q-gates} & \textbf{2q-gates} & \textbf{Depth} \\ \midrule
        \multirow{2}{*}{Unencoded} & All-to-all & 24  & 16  & 28 \\
                                   & Routed     & 24  & 34  & 36 \\ \midrule
        \multirow{2}{*}{Encoded}   & All-to-all & 130 & 172 & 137 \\
                                   & Routed     & 130 & 448 & 227 \\ \bottomrule
    \end{tabular}
\end{table}

\subsection{Noise Model}
To assess the performance of our logical implementation, we characterized the post-selection rate and the ground state probability of post-selected samples across the following noise components:

\begin{itemize}
    \item \textbf{Single-qubit gate error}: after each single-qubit gate, a depolarizing channel is applied, introducing a Pauli $X$, $Y$, or $Z$ error with probability $p$.

    \item \textbf{Two-qubit gate error}: after each two-qubit gate, a random two-qubit Pauli error (e.g., $XX$, $XY$, $IX$, $XI$) is applied with probability $p$.

    \item \textbf{Amplitude damping}: after each single- and two-qubit unitary gate, an amplitude damping channel with decay probability $p$ is applied to all qubits involved in order to model energy relaxation.

    \item \textbf{Readout error}: a bit-flip is applied with probability $p$ immediately before each qubit measurement.

    \item \textbf{Reset error}: a bit-flip is applied with probability $p$ immediately after each qubit reset.
\end{itemize}

\subsection{Multithreading vs.\ GPU Simulation}
Circuits were constructed using the Google Cirq framework \cite{Cirq_2025}, and simulations were performed with the \texttt{qsimcirq} backend \cite{qsim}. 
The complete source code used to obtain the results presented in this paper is publicly available~\cite{codeRepo}. 
The CPU used was an AMD EPYC 9354 running at 3250\,MHz; the GPU was an NVIDIA L40S. GPU-based simulations employed NVIDIA cuQuantum with CUDA (\texttt{gpu\_mode=0}). For the all-to-all encoded circuit under a minimal noise model comprising only two-qubit depolarizing errors at $p = 0.01$, a single CPU thread was approximately $3\times$ faster than the GPU. 
We attribute this to the overhead of transferring the circuit to and from GPU memory, which outweighs the computational speedup that the GPU would otherwise provide for circuits of this size.

\begin{figure}
    \centering
    \includegraphics[width=1.\linewidth]{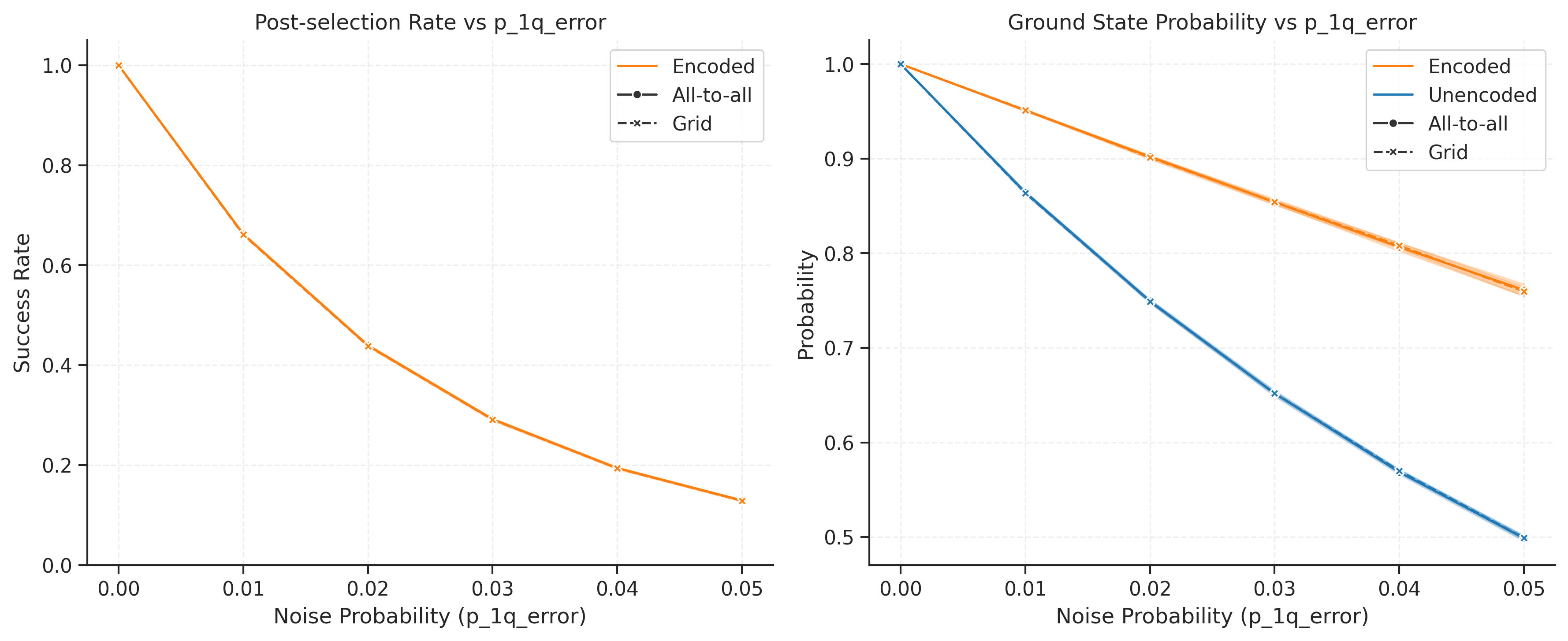}
    \includegraphics[width=1.\linewidth]{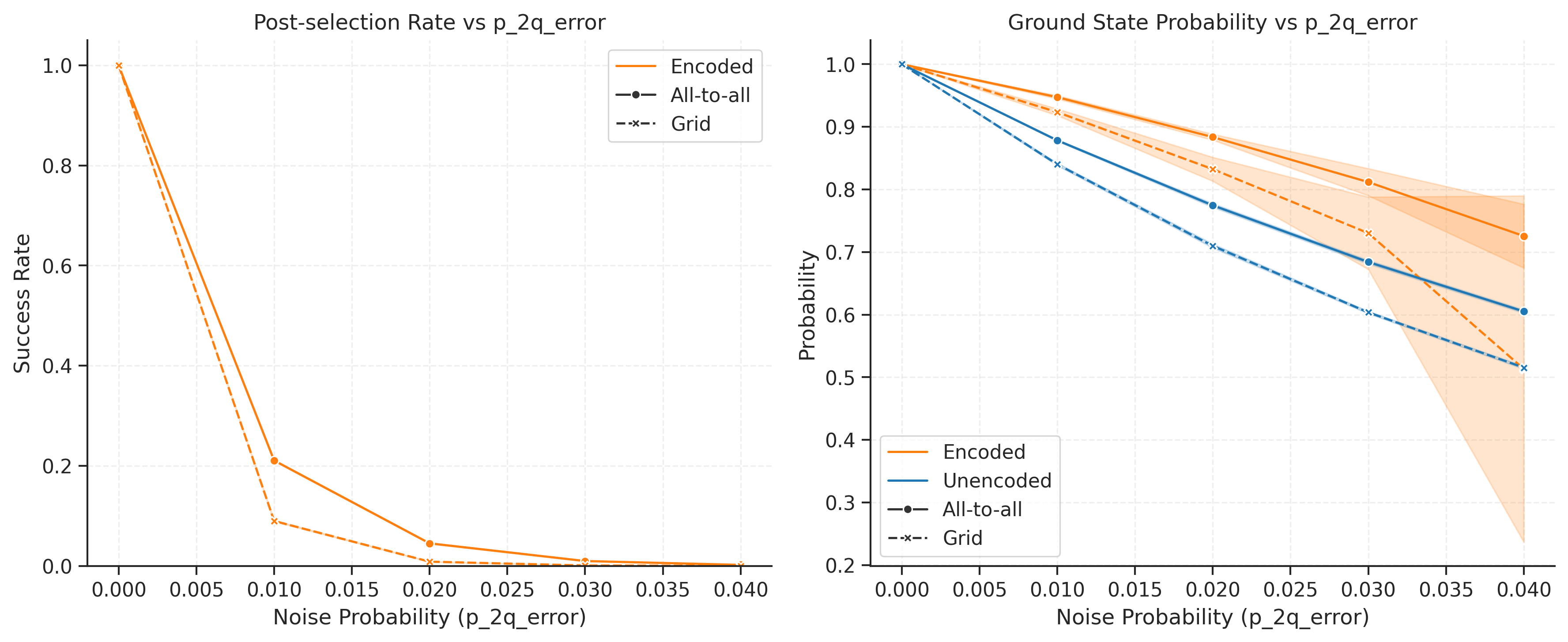}
    \caption{Post-selection rate (left) and ground state probability of post-selected samples (right) as a function of noise strength $p$, for single-qubit (top) and two-qubit (bottom) depolarizing noise. Line colour denotes encoding: orange for the [[4,2,2]]-encoded circuit, blue for the unencoded baseline. Line style denotes topology: solid for the all-to-all connectivity, dashed for the grid-routed circuit. Shaded bands represent $\pm1$ standard deviation across the 10 experimental runs.}
    \label{fig:1q2q-errors}
\end{figure}

\subsection{Post-selection Rate and Ground State Probability}
We conducted 10 experiments with 50{,}000 repetitions each across 24 different noise configurations, for a total of 240 experiments, with each experiment executed in parallel on a dedicated CPU thread. Across all graphs, error bars for the post-selection rate are not visible due to their small magnitude.

The results vary qualitatively depending on the noise model. Under \textit{single-qubit depolarizing noise}, the post-selection rate and ground state probability curves for the all-to-all and routed unencoded circuits overlap, since routing introduces only additional two-qubit gates, which are unaffected by this noise model. The encoded circuit, however, achieves a markedly higher ground state probability among post-selected samples. This can be seen in the top row of Figure~\ref{fig:1q2q-errors}.

Under \textit{two-qubit depolarizing noise} (bottom row of Figure~\ref{fig:1q2q-errors}), routing the circuit onto the grid substantially reduces the post-selection rate, as the additional SWAP gates introduce more opportunities for error. Despite this, the encoded routed circuit still improves ground state probability relative to its unencoded counterpart for noise levels up to $p = 0.03$, while the all-to-all encoded circuit maintains this advantage up to $p = 0.04$. The error bars for both encoded circuits grow notably with the noise probability, reflecting the reduced number of post-selected samples from which the standard deviation is estimated.

\begin{figure}
    \centering
    \includegraphics[width=1.\linewidth]{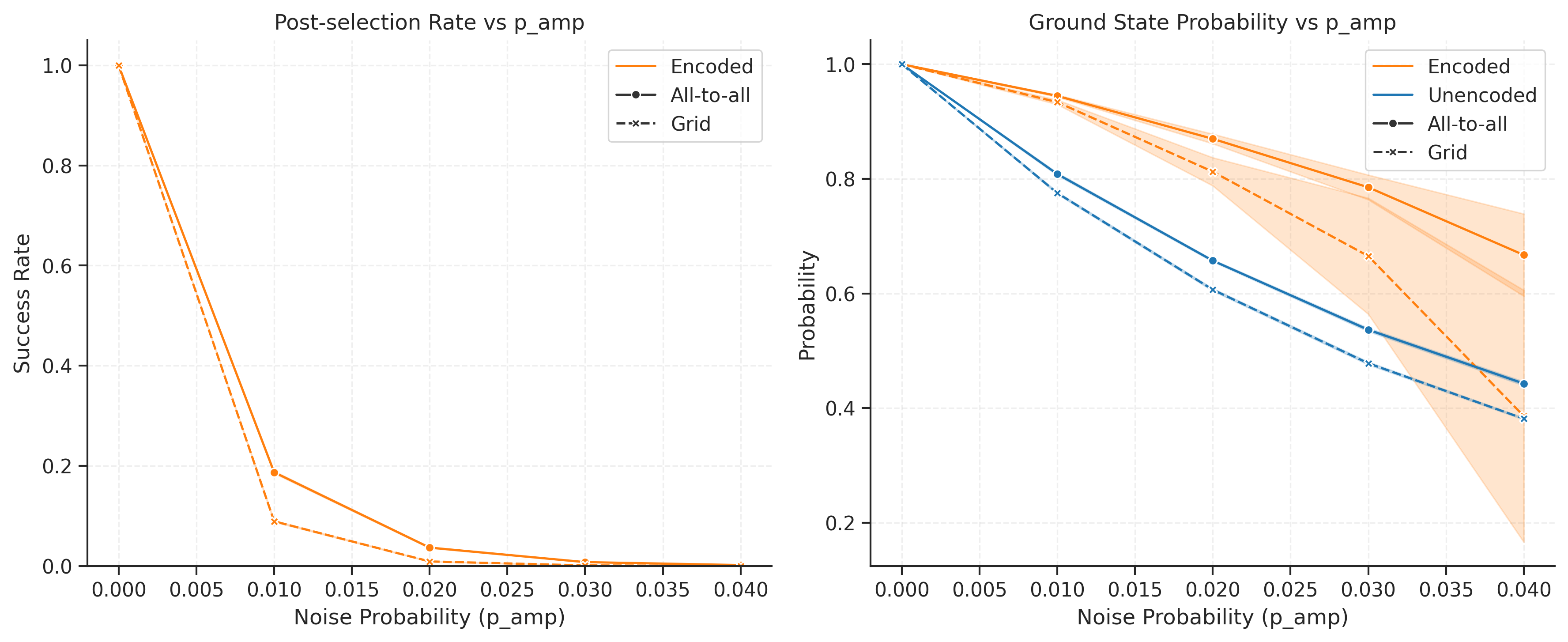}
    \includegraphics[width=1.\linewidth]{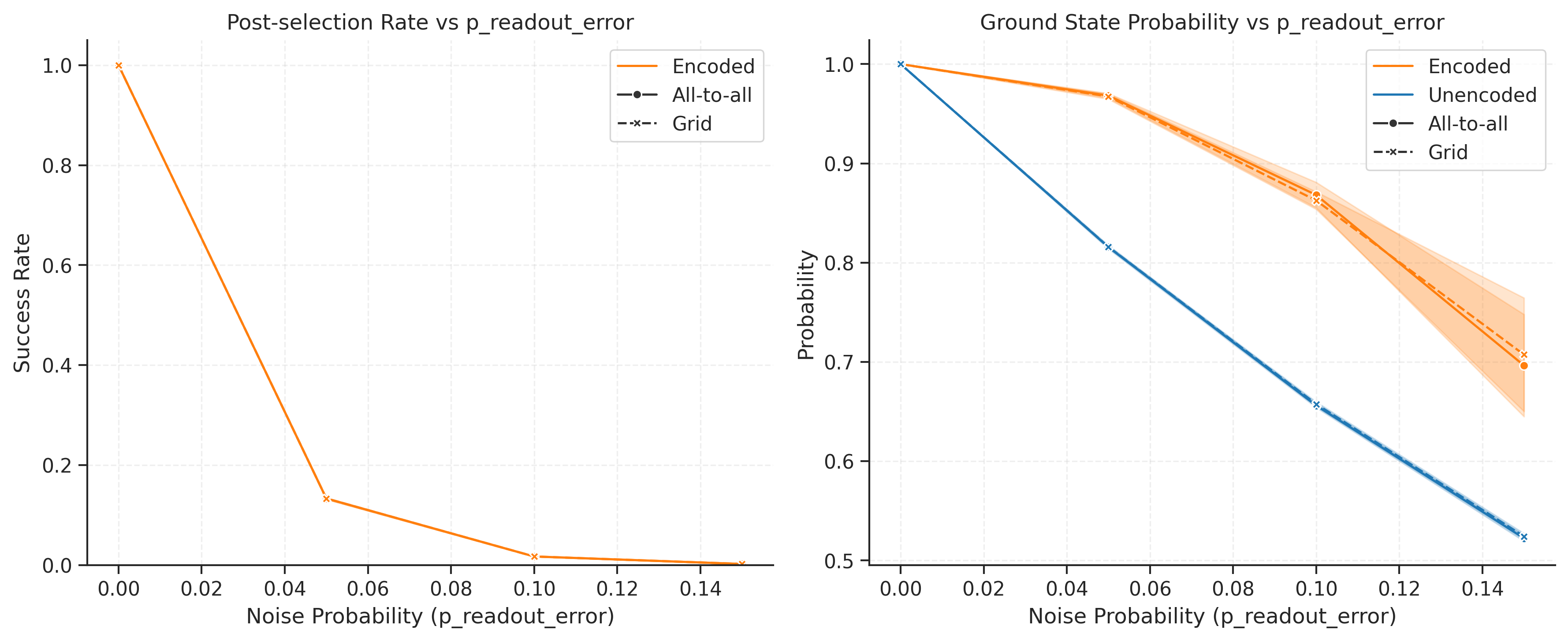}
    \includegraphics[width=1.\linewidth]{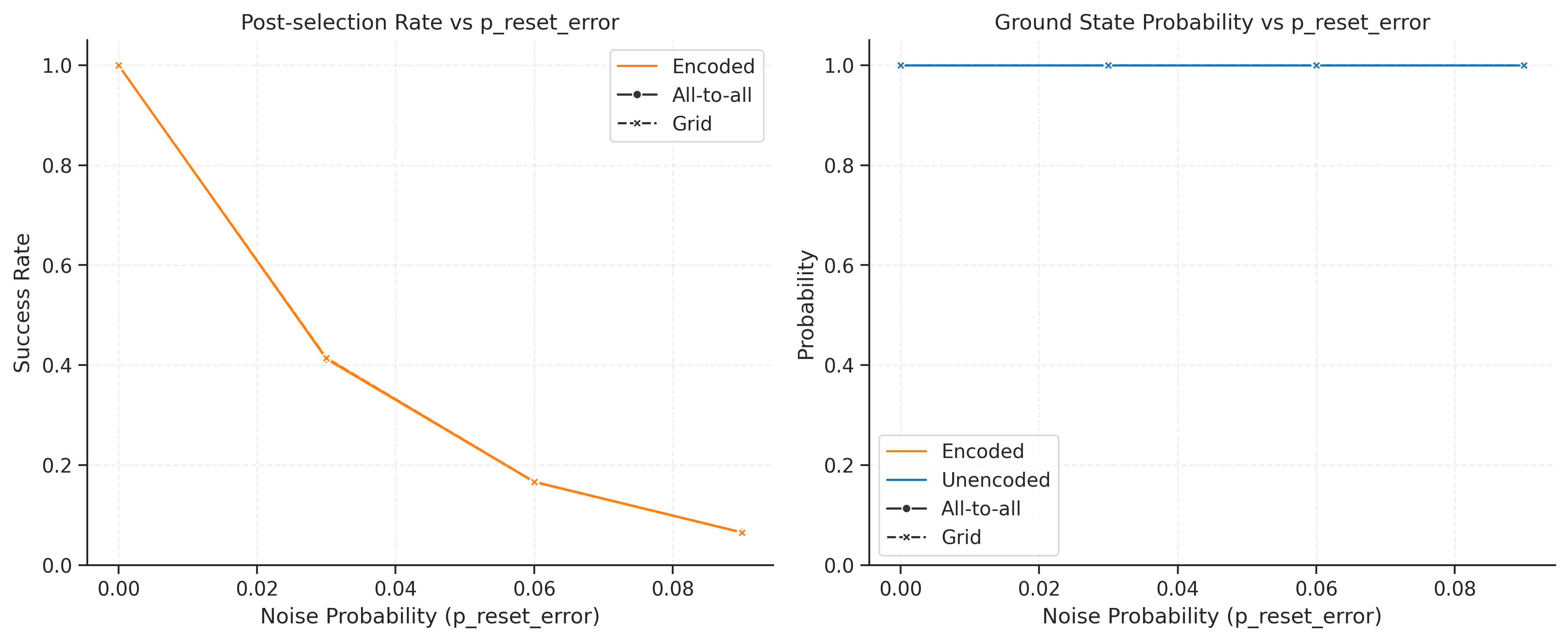}
    \caption{Post-selection rate (left) and ground state probability of post-selected samples (right) as a function of noise strength $p$, for amplitude damping (top), readout error (middle), and reset error (bottom). Line colour denotes encoding: orange for the [[4,2,2]]-encoded circuit, blue for the unencoded baseline. Line style denotes topology: solid for the all-to-all connectivity, dashed for the grid-routed circuit. Shaded bands represent $\pm1$ standard deviation across the 10 experimental runs.}
    \label{fig:other-errors}
\end{figure}

\textit{Amplitude damping} yields a qualitatively similar picture, showcased in the first row of Figure~\ref{fig:other-errors}. The encoded circuits outperform their unencoded counterparts at least up to $p = 0.03$. Under \textit{readout error} (middle row of Figure~\ref{fig:other-errors}), the post-selection rate and ground state probability curves for the routed and all-to-all variants of each circuit coincide precisely, since the embedding does not affect measurement operations. A small divergence appears in the ground state probability at higher noise levels, where the reduced number of post-selected samples introduces greater statistical uncertainty.

Finally, under \textit{reset error} (bottom row of Figure~\ref{fig:other-errors}), the ground state probability of post-selected samples is exactly 1 for all circuits, regardless of encoding or topology. This is a direct consequence of the error-detection mechanism: if an ancilla qubit is incorrectly initialised to $|1\rangle$ instead of $|0\rangle$ and no other errors occur, it will also be measured as $|1\rangle$. It will therefore be discarded by the post-selection filter, ensuring that only correctly initialised runs contribute to the statistics.

\section{Discussion \& Conclusions}
\label{sec:conclusions}

We have presented a partially fault-tolerant implementation of QAOA for the Max-Cut problem on a four-vertex square graph, based on the $[[4,2,2]]$ quantum error-detection code.
Our main technical contribution is an ancilla-mediated construction of logical $R_{ZZ}$ gates acting between distinct $[[4,2,2]]$ blocks, extending prior work that only considered intra-block rotations.
Evaluation under five noise channels shows that the encoded circuit consistently achieves higher ground-state probability among post-selected samples than its unencoded counterpart. 
The approach fits naturally into a quantum--HPC hybrid workflow: parameter concentration allows near-optimal QAOA angles to be computed offline on small instances (trivially parallelisable across HPC workers) and then deployed on the quantum device with minimal retraining.
As an anecdotal observation, CPU-based simulation with many parallel threads outperformed the GPU-backed backend for circuits of this size; a more systematic sweep over circuit widths, depths, and batch sizes is left for future work.

Several directions remain open.
The classical cost of syndrome decoding should be quantified for full resource accounting at scale.
Circuit depth, roughly $5\times$ that of the unencoded circuit, could be reduced by using additional ancilla qubits to parallelise stabilizer rounds and logical rotations.
Scaling to larger Max-Cut instances will require balancing the benefits of error suppression against the post-selection overhead on deeper circuits. We stress that our evaluation is confined to a single $C_4$ instance spanning two blocks: while the inter-block $R_{ZZ}$ construction is by design sufficient to encode arbitrary instances across more blocks, the net advantage of post-selection at larger sizes remains to be established empirically. In fact, the post-selection rate compounds multiplicatively with circuit depth and the number of stabilizer rounds.
Finally, transitioning from error detection to full error correction would eliminate the sampling overhead entirely, at the cost of a substantially larger qubit count, the natural next step toward production-grade fault-tolerant quantum--HPC workflows.

\begin{credits}

\subsubsection{\discintname}
The authors have no competing interests to declare that are
relevant to the content of this article.
\end{credits}
%
%
%
\bibliographystyle{splncs04}
\bibliography{ref.bib}

@misc{bedalovFaultTolerantOperationMaterials2024,
  title = {Fault-{{Tolerant Operation}} and {{Materials Science}} with {{Neutral Atom Logical Qubits}}},
  author = {Bedalov, Matt J. and others},
  year = 2024,
  month = dec,
  number = {arXiv:2412.07670},
  eprint = {2412.07670},
  primaryclass = {quant-ph},
  publisher = {arXiv},
  doi = {10.48550/arXiv.2412.07670},
  urldate = {2025-12-10},
  archiveprefix = {arXiv},
  langid = {english}
}

@article{hePerformanceQuantumApproximate2025,
  title = {Performance of {{Quantum Approximate Optimization}} with {{Quantum Error Detection}}},
  author = {He, Zichang and Amaro, David and Shaydulin, Ruslan and Pistoia, Marco},
  year = 2025,
  month = may,
  journal = {Communications Physics},
  volume = {8},
  number = {1},
  eprint = {2409.12104},
  primaryclass = {quant-ph},
  pages = {217},
  issn = {2399-3650},
  doi = {10.1038/s42005-025-02136-8},
  urldate = {2025-12-10},
  archiveprefix = {arXiv},
  langid = {english}
}

@misc{Farhi2014QAOA,
  doi = {10.48550/ARXIV.1411.4028},
  url = {https://arxiv.org/abs/1411.4028},
  author = {Farhi,  Edward and Goldstone,  Jeffrey and Gutmann,  Sam},
  title = {A Quantum Approximate Optimization Algorithm},
  publisher = {arXiv},
  year = {2014},
  copyright = {arXiv.org perpetual,  non-exclusive license}
}

@article{cerezo2023simulability,
  title = {Does provable absence of barren plateaus imply classical simulability?},
  volume = {16},
  ISSN = {2041-1723},
  url = {http://dx.doi.org/10.1038/s41467-025-63099-6},
  DOI = {10.1038/s41467-025-63099-6},
  number = {1},
  journal = {Nature Communications},
  publisher = {Springer Science and Business Media LLC},
  author = {Cerezo,  M. and Larocca,  Martin and García-Martín,  Diego and Diaz,  N. L. and Braccia,  Paolo and Fontana,  Enrico and Rudolph,  Manuel S. and Bermejo,  Pablo and Ijaz,  Aroosa and Thanasilp,  Supanut and Anschuetz,  Eric R. and Holmes,  Zoë},
  year = {2025},
  month = Aug 
}

@article{cerezo2021variational,
  title = {Variational quantum algorithms},
  volume = {3},
  ISSN = {2522-5820},
  url = {http://dx.doi.org/10.1038/s42254-021-00348-9},
  DOI = {10.1038/s42254-021-00348-9},
  number = {9},
  journal = {Nature Reviews Physics},
  publisher = {Springer Science and Business Media LLC},
  author = {Cerezo,  M. and Arrasmith,  Andrew and Babbush,  Ryan and Benjamin,  Simon C. and Endo,  Suguru and Fujii,  Keisuke and McClean,  Jarrod R. and Mitarai,  Kosuke and Yuan,  Xiao and Cincio,  Lukasz and Coles,  Patrick J.},
  year = {2021},
  month = Aug,
  pages = {625–644}
}

@misc{farhiHarrow2016supremacy,
  author = {Farhi, Edward and Harrow, Aram W.},
  title = {Quantum Supremacy through the Quantum Approximate 
           Optimization Algorithm},
  year = {2019},
  eprint = {1602.07674},
  archivePrefix = {arXiv},
  primaryClass = {quant-ph},
  doi = {10.48550/arXiv.1602.07674}
}

@article{mcclean2018barren,
  title = {Barren plateaus in quantum neural network training landscapes},
  volume = {9},
  ISSN = {2041-1723},
  url = {http://dx.doi.org/10.1038/s41467-018-07090-4},
  DOI = {10.1038/s41467-018-07090-4},
  number = {1},
  journal = {Nature Communications},
  publisher = {Springer Science and Business Media LLC},
  author = {McClean,  Jarrod R. and Boixo,  Sergio and Smelyanskiy,  Vadim N. and Babbush,  Ryan and Neven,  Hartmut},
  year = {2018},
  month = Nov 
}

@article{Akshay2021,
  title = {Parameter concentrations in quantum approximate optimization},
  volume = {104},
  ISSN = {2469-9934},
  url = {http://dx.doi.org/10.1103/PhysRevA.104.L010401},
  DOI = {10.1103/physreva.104.l010401},
  number = {1},
  journal = {Physical Review A},
  publisher = {American Physical Society (APS)},
  author = {Akshay,  V. and Rabinovich,  D. and Campos,  E. and Biamonte,  J.},
  year = {2021},
  month = Jul
}

@article{wurtz2023sampling,
  title = {Sampling frequency thresholds for the quantum advantage of the quantum approximate optimization algorithm},
  volume = {9},
  ISSN = {2056-6387},
  DOI = {10.1038/s41534-023-00718-4},
  number = {1},
  journal = {npj Quantum Information},
  publisher = {Springer Science and Business Media LLC},
  author = {Lykov,  Danylo and Wurtz,  Jonathan and Poole,  Cody and Saffman,  Mark and Noel,  Tom and Alexeev,  Yuri},
  year = {2023},
  month = Jul
}

@article{hao2025endtoend,
  title = {End-to-end protocol for high-quality quantum approximate optimization algorithm parameters with few shots},
  volume = {7},
  ISSN = {2643-1564},
  url = {http://dx.doi.org/10.1103/24gg-7p8z},
  DOI = {10.1103/24gg-7p8z},
  number = {3},
  journal = {Physical Review Research},
  publisher = {American Physical Society (APS)},
  author = {Hao,  Tianyi and He,  Zichang and Shaydulin,  Ruslan and Larson,  Jeffrey and Pistoia,  Marco},
  year = {2025},
  month = Aug 
}

@article{Goemans1995,
  title = {Improved approximation algorithms for maximum cut and satisfiability problems using semidefinite programming},
  volume = {42},
  ISSN = {1557-735X},
  url = {http://dx.doi.org/10.1145/227683.227684},
  DOI = {10.1145/227683.227684},
  number = {6},
  journal = {Journal of the ACM},
  publisher = {Association for Computing Machinery (ACM)},
  author = {Goemans,  Michel X. and Williamson,  David P.},
  year = {1995},
  month = Nov,
  pages = {1115–1145}
}

@misc{Cirq_2025,
  author       = {{Cirq Developers}},
  title        = {Cirq},
  year         = {2025},
  howpublished = {\url{https://zenodo.org/doi/10.5281/zenodo.4062499}}
}

@misc{qsim,
  author       = {{Quantum AI team}},
  title        = {qsim},
  year         = {2025},
  howpublished = {\url{https://doi.org/10.5281/zenodo.4067237}}
}

@misc{codeRepo,
  author       = {{Child, Matteo R.}},
  title        = {Replication code},
  year         = {2026},
  url          = {https://github.com/matchild/logical-qaoa}
}

@article{Reichardt2020,
  title = {Fault-tolerant quantum error correction for Steane’s seven-qubit color code with few or no extra qubits},
  volume = {6},
  ISSN = {2058-9565},
  url = {http://dx.doi.org/10.1088/2058-9565/abc6f4},
  DOI = {10.1088/2058-9565/abc6f4},
  number = {1},
  journal = {Quantum Science and Technology},
  publisher = {IOP Publishing},
  author = {Reichardt,  Ben W},
  year = {2020},
  month = Nov,
  pages = {015007}
}

@article{Vaidman1996,
  title = {Error prevention scheme with four particles},
  volume = {54},
  ISSN = {1094-1622},
  url = {http://dx.doi.org/10.1103/PhysRevA.54.R1745},
  DOI = {10.1103/physreva.54.r1745},
  number = {3},
  journal = {Physical Review A},
  publisher = {American Physical Society (APS)},
  author = {Vaidman,  Lev and Goldenberg,  Lior and Wiesner,  Stephen},
  year = {1996},
  month = Sep,
  pages = {R1745–R1748}
}

@article{Grassl1997,
  title = {Codes for the quantum erasure channel},
  volume = {56},
  ISSN = {1094-1622},
  url = {http://dx.doi.org/10.1103/PhysRevA.56.33},
  DOI = {10.1103/physreva.56.33},
  number = {1},
  journal = {Physical Review A},
  publisher = {American Physical Society (APS)},
  author = {Grassl,  M. and Beth,  Th. and Pellizzari,  T.},
  year = {1997},
  month = Jul,
  pages = {33–38}
}
%




\end{document}